\documentclass[reprint, amsmath,amssymb,aps,prb,superscriptaddress]{revtex4-2}

\usepackage{graphicx}
\usepackage{dcolumn}
\usepackage{bm}
\usepackage[usenames,dvipsnames]{color}

\begin{document}

\title{Linear scaling relationship of N\'{e}el temperature and dominant magnons in pyrochlore ruthenates}
\author{Jae Hyuck Lee}\thanks{Both authors equally contributed to this work}\affiliation{Center for Correlated Electron Systems, Institute for Basic Science, Seoul 08826, Republic of Korea}\affiliation{Department of Physics and Astronomy, Seoul National University, Seoul 08826, Republic of Korea} 
\author{Dirk Wulferding}\thanks{Both authors equally contributed to this work}\affiliation{Center for Correlated Electron Systems, Institute for Basic Science, Seoul 08826, Republic of Korea}\affiliation{Department of Physics and Astronomy, Seoul National University, Seoul 08826, Republic of Korea} 
\author{Junkyoung Kim} \affiliation{Department of Physics, Incheon National University, Incheon 22012, Republic of Korea}
\author{Dongjoon Song}\email[E-mail: ]{scdjsong@gmail.com}\affiliation{Center for Correlated Electron Systems, Institute for Basic Science, Seoul 08826, Republic of Korea}\affiliation{Department of Physics and Astronomy, Seoul National University, Seoul 08826, Republic of Korea}\affiliation{Stewart Blusson Quantum Matter Institute, University of British Columbia, Vancouver, British Columbia V6T 1Z4, Canada}
\author{Seung Ryong Park}\email[E-mail: ]{abepark@inu.ac.kr}\affiliation{Department of Physics, Incheon National University, Incheon 22012, Republic of Korea}
\author{Changyoung Kim}\affiliation{Center for Correlated Electron Systems, Institute for Basic Science, Seoul 08826, Republic of Korea}\affiliation{Department of Physics and Astronomy, Seoul National University, Seoul 08826, Republic of Korea} 
\email[REVTeX Support: ]{revtex@aps.org}
\date{\today}
\begin{abstract}
We present a systematic Raman spectroscopy study on a series of pyrochlore ruthenates, a system which is not yet clearly settled on its magnetic origin and structure. Apart from the Raman-active phonon modes, new peaks that appear in the energy range of 15 - 35 meV below the N\'{e}el temperature (\textit{T}\textsubscript{N}) are assigned as one-magnon modes. The temperature evolution of one-magnon modes displays no significant thermal dependence in mode frequencies while the intensities decrease monotonically. Remarkably, one-magnons from all compounds show similar characteristics with a single dominant peak at lower energy and weaker side peaks at a couple of meV higher energy. Most importantly, we uncover a striking proportionality between the dominant magnon mode energies and the \textit{T}\textsubscript{N} values. Our results suggest the Ru ions may have similar or the same magnetic phase in all pyrochlore ruthenates of our study. We have thus found an avenue for directly tuning the magnetic exchange interaction by the selection of the $A$-site ion.
\end{abstract}
\maketitle
\section{Introduction}

Pyrochlore ruthenates, a group of $4d$ transition metal oxides (TMO), are natural candidates for exploring tunable magnon band structures, as their intrinsic competing energy scales can lead to the formation of a wide range of magnetic ground states based on the details of the system. Pyrochlore oxides with the chemical formula $A_2$$B_2$O$_7$ ($A$= Pr-Lu, Y and $B$=Ru for pyrochlore ruthenates) crystallize in the cubic space group $Fd\bar{3}m$ built upon two interpenetrating sublattices of corner-shared tetrahedra. The $A$ and $B$-site ions form the corners of the tetrahedra, while 6 oxygen ions surround the $B$-site forming a connected network of $B$O$_6$ octahedra, and 8 oxygen ions around the $A$-site having strong axial symmetry where the easy axis is along the [111] direction~\cite{Rev1}.

The tetrahedral nature of this 3-dimensional lattice inherently hosts geometric magnetic frustration, making it the perfect playground for exploring and tuning novel magnetic states such as spin ice~\cite{spinice1,spinice3, spinice5,spinice6,spinice7,spinice9,spinice10}, spin glass~\cite{spinglass2, spinglass4, spinglass5, spinglass6}, as well as quantum spin liquids~\cite{spinglass5,spinliquid1,  spinliquid3, spinliquid5,spinliquid6,spinliquid7, spinliquid10,spinliquid11,spinice8}. Materials with pyrochlore structure also form kagome lattices stacked along the [111] direction, which can realize highly sought-after flat-band structures as well as linear bands with Dirac points, i.e., non-trivial band topologies~\cite{kagome1,PyroFlat1,PyroFlat2,PyroFlat3,PyroFlat4}.

In general, it has been reported that pyrochlore ruthenates experience two distinct magnetic phase transitions; one is at higher temperature and attributed to the Ru spins while the other is at a much lower temperature and related to ordering of the $A$-site rare earth (RE) ions. Here, we focus on high temperature magnetic transitions driven largely by the Ru ion on the $B$-site. However, a fundamental question on the magnetism remains regarding the exact magnetic origin from the $t_{2g}^4$ Ru$^{4+}$ ions. When considering only the spins, by Hund's coupling the ions realize a local low-spin state, therefore we can classify them as S=1 pyrochlore magnets~\cite{Gao}. On the other hand, introducing substantial spin-orbit coupling results in excitonic magnetism in which the J=1 state is stabilized via inter-atomic exchange coupling at low temperature, otherwise a non-magnetic J=0 state exists at high temperature~\cite{kaulExiMag}. 

Moreover, the magnetic structure of these materials has not been pinned down conclusively~\cite{Gao}. As with most $4d$ TMOs, pyrochlore ruthenates are conflicted by strongly competing spin-orbit coupling and crystal field splitting, resulting in an oftentimes ill-defined ground state~\cite{Rev1,Gao}. While in general the Ru ions undergo transitions into an ordered state at the N\'{e}el temperature (\textit{T}\textsubscript{N}) due to the dominant Ru-Ru exchange interactions, $A$-Ru exchange interactions as well as $A$-$A$ correlations may be non-negligible~\cite{Gao,Rev1} and the exact spin structures are still under debate for many of the compounds~\cite{Gao,Rev1,mag1,mag2,mag3Er,mag6Er,mag7HoDy,mag71Ho,mag8Ho,mag9Ho,Gd227,mag11Y,mag12Eu,mag13Eu,mag14Nd,mag15Nd,mag16Nd,mag17Nd,mag17Nd2,mag18}. 

In elucidating the magnetic origin and spin structure in $A_2$Ru$_2$O$_7$, the magnon band structure would give us information about such magnetic properties. Magnon modes can be one of the most direct guidelines in determining magnetic exchange interaction (J) or Dzyaloshinskii-Moriya interaction (D) and the spin structure. For a system that holds ambiguity over the nature of its magnetic phases, such observations are critical. Indeed, temperature dependent Raman spectroscopy on pyrochlore iridates displayed emergence of one-magnon excitations below \textit{T}\textsubscript{N} and accurate values of the ratio J/D were determined~\cite{YIORaman}.
Here we aim to extend such an approach to pyrochlore ruthenates and investigate the behavior of such excitations in this family of compounds~\cite{YIORaman, LF, ELF,RamanMag, RamanMag1,RamanMag3,RamanMag4,RamanMag6,RamanMag8,RamanMag9,RamanMag10,RamanMag11,RamanMag12,RamanMag13,RamanMag14,NDSC}. 

We conduct a systematic approach via Raman spectroscopy by varying the chemical pressure through changing the $A$-site ion ($A$= Y, Nd, Sm, Eu, Ho, and Er). Raman scattering is an experimental tool with high spectral resolution suited to probe magnetic excitations at q $\approx$ 0 in strongly correlated materials~\cite{RamanMag3,RamanMag4,RamanMag6,RamanMag8,RamanMag9,RamanMag10,RamanMag11,RamanMag12,RamanMag13,RamanMag14,EIOmagnon,RamanMag1,SIOmagnon1,SIOmagnon2,Ruclmagnon0,RamanMag9}. Our study reveals that for all compounds, one dominant magnon peak and a couple of additional side peaks appear only below \textit{T}\textsubscript{N}. Remarkably, the energy of the dominant magnon scales proportionally with \textit{T}\textsubscript{N}. This suggests a common magnetic ground state for this pyrochlore ruthenate family. By comparing magnetic properties across a wide range of compounds we pave a straightforward avenue for tuning the magnetic band structure in $4d$ TMO pyrochlore ruthenates.
\section{Experimental Details}
\textit{Sample synthesis.} A series of polycrystalline samples was synthesized through standard solid state reaction: Stoichiometric quantities of starting materials consisting of RuO$_2$ and preheated $A_2$O$_3$ ($A$= Y, Nd, Sm, Eu, Ho, and Er) were mixed together with an excessive 1 mol\% RuO$_2$ to compensate for the high volatility of ruthenium oxide. The ingredients were thoroughly mixed using an agate mortar and pestle for about an hour.  The fine powders were then put into alumina crucibles and sintered in air at 1173 K - 1573 K for approximately 48 hours with intermediate grindings.

\textit{Sample characterization.} Pelletized samples are characterized by powder x-ray diffraction (p-XRD). XRD patterns were obtained with D8 Advanced (Bruker Inc, Germany) for all $A_2$Ru$_2$O$_7$ compounds. The data was measured with an angle increment of 0.02$^{\circ}$. dc-magnetic susceptibility data were measured using a superconducting quantum interference device magnetometer (Quantum Design MPMS3 model). Zero-field-cooled (ZFC) and field-cooled (FC) temperature dependent dc-magnetic susceptibility curves were obtained at a field of 100 Oe.

\textit{Raman spectroscopy.} Raman spectroscopic measurements were performed using a home-built setup. A ${\lambda}$ = 532 nm solid state laser (Cobolt SambaTM-500 series) was used for excitation. A 40$\times$ microscope objective (Olympus) focused the laser onto the samples with a spot diameter of 6 $\mu$m. The laser power was set to 0.05 mW for $A$=Y, Er, and to 0.1 mW for $A$=Sm, Eu, Ho. Raman scattered light passed through a set of 3 notch filters (Optigrate) to suppress the laser line. The spectrometer (iHR-320 with Horiba Synapse CCD) was equipped with a 1800 gr/mm grating. Measurements were carried out with 900 seconds acquisition time and 10 accumulations. While both parallel and crossed polarization were conducted, a parallel light polarization configuration is chosen for all results displayed in the figures [see Supplemental Material Fig. S1~\cite{suppl}].

\section{Results and Discussion}
\begin{figure*}
    \includegraphics[width=18cm]{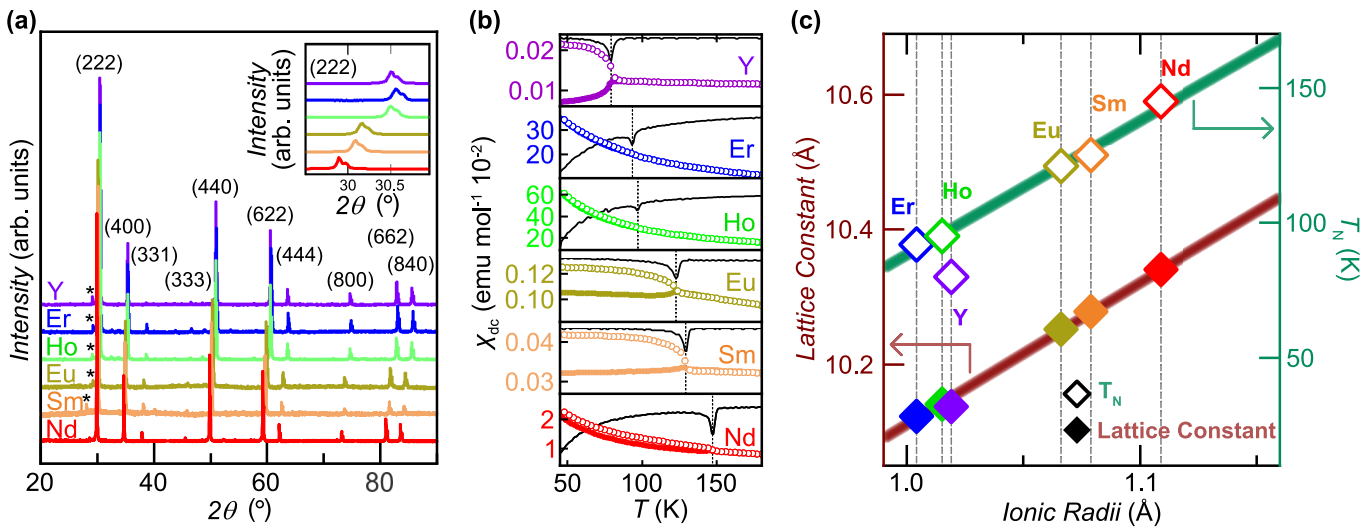}
    \caption{\textbf{Basic Characterization of $A_2$Ru$_2$O$_7$.} (a) p-XRD measurements of $A_2$Ru$_2$O$_7$ taken at room temperature. Assigned Bragg peaks are labelled. Asterisks denote an impurity phase. The inset zooms into the main (222) peak where the double peaked feature comes from the Cu-$K$$_\alpha$$_1$ and Cu-$K$$_\alpha$$_2$. (b) Temperature-dependent magnetic susceptibility ($\chi$\textsubscript{dc}) curves of $A_2$Ru$_2$O$_7$. Filled circles: ZFC; open circles: FC at B=100 Oe. Black solid lines represent $\frac{\mathrm{d}\chi}{\mathrm{d}T}$ of the FC curves. The kinks of the first derivatives and dotted vertical lines indicate \textit{T}\textsubscript{N}. (c) Lattice constant and \textit{T}\textsubscript{N} dependence on the $A$-site RE ion size. Filled diamonds represent the lattice constants calculated from the p-XRD data shown in (a), and the open diamonds stand for the \textit{T}\textsubscript{N} obtained from temperature-dependent magnetization data of (b). Thick solid lines for lattice constant and \textit{T}\textsubscript{N} are indicated with the brown left axis representing the lattice constant and the green right axis representing the \textit{T}\textsubscript{N} as directed by the arrows.}
    \label{fig1}
\end{figure*}

In order to confirm the pyrochlore structure and to rule out secondary phases, we first turn to p-XRD results shown in Fig. \ref{fig1}(a). Except for a minute amount of RE oxide phases $A_2$O$_3$ [marked by the asterisks in Fig. \ref{fig1}(a)], no other secondary phase was detected. Based on the p-XRD results, we can assign the structure as cubic $Fd\bar{3}m$ as expected for pyrochlore lattices~\cite{Rev1}. The main Bragg peaks of each compound corresponding to (222) orientation are shown in detail in the inset of Fig. \ref{fig1}(a), emphasizing a shift of Bragg peaks with respect to the $A$-site atom. From those (222) peaks, we have calculated the lattice constants for all compounds, which are plotted with respect to the ionic radii of the $A$-site ions in Fig. \ref{fig1}(c). Note that between the $A$-site ionic radius and lattice constant, there is a linear relation in agreement with an empirical analysis ~\cite{pyrochloreLatticeCons}. The brown fitting line in Fig. \ref{fig1}(c) matches considerably well with the model equation (3) given in Ref~\cite{pyrochloreLatticeCons}: a $\approx$ 2 $R_A$ + 8 \AA, where a (\AA) and $R_A$ (\AA) stand for the lattice constant and ionic radius of the $A$-site RE ion respectively.

Next, we characterize the magnetic properties via magnetic susceptibility data shown in Fig. \ref{fig1}(b). The ZFC and FC measurement of the synthesized compounds are each given as filled and open circles, respectively. A branching of the ZFC and FC magnetic susceptibility below \textit{T}\textsubscript{N} is found in all compounds although the extent of the bifurcation varies depending on the $A$-site ion. This is common for geometrically frustrated systems~\cite{doublepero}. Moreover, branching between ZFC and FC can also indicate hysteresis or spin glass behavior. However, it was suggested in neutron scattering, heat capacity measurements and ac susceptibility experiments that below \textit{T}\textsubscript{N} Ru spins order while displaying no frequency dependence of susceptibility, inconsistent with a spin glass phase~\cite{mag3Er,mag9Ho,mag14Nd,mag17Nd,mag17Nd2}. Above \textit{T}\textsubscript{N}, the magnetic susceptibility curves for all compounds follow a Curie-Weiss behavior [$\chi=C/(T-\Theta_{CW})$, with $C$ being the Curie constant], yielding negative Weiss temperatures $\Theta_{CW}$. This clearly indicates antiferromagnetic interactions between the Ru ions. 
Our results are consistent with previous studies conducted systematically or individually on this family of compounds~\cite{Gao,Rev1,mag1,mag2,mag3Er,mag6Er,mag7HoDy,mag71Ho,mag8Ho,mag9Ho,Gd227,mag11Y,mag12Eu,mag13Eu,mag14Nd,mag15Nd,mag16Nd,mag17Nd,mag18}. Note that there is a Curie-tail type increase of the susceptibility below \textit{T}\textsubscript{N}, contrary to typical antiferromagnetic ordering behavior, which can be mainly attributed to paramagnetic impurities from the $f$-electrons.

\textit{T}\textsubscript{N} for each compound are marked by black arrows in Fig. \ref{fig1}(b). Note that the value of \textit{T}\textsubscript{N} varies depending on the RE ions on the $A$-site. In fact, \textit{T}\textsubscript{N} displays a linear relation with the ionic radius of the $A$-site ions as shown in Fig. \ref{fig1}(c), with Y being the only outlier. In contrast to other $A$-site elements, Y lacks a filled $4f$ electron state. Recalling that the lattice constant also linearly depends on the $A$-site ionic radius [see brown line in Fig. \ref{fig1}(c)], \textit{T}\textsubscript{N} and the lattice constant must also be linearly correlated. It was reported that increasing the $A$-site ionic radius increases the Ru-O-Ru angle in the $B$-site sublattice RuO$_6$ octahedra and that they follow a linear relationship~\cite{whangbo}. Negative chemical pressure due to a smaller ionic radius at the $A$-site leads to a compressive distortion of the RuO$_6$ octahedra. This in turn decreases the Ru-O-Ru angles, together with the hopping amplitude ($t$) and the bandwidth ($W$) of the $t_{2g}$ orbital band of the Ru $4d$ electrons, and thereby also the superexchange interaction (J) via J $\propto$ $t^2$/$U$~\cite{goodenough, tokura}.

\begin{figure}
    \includegraphics[width=8.5cm]{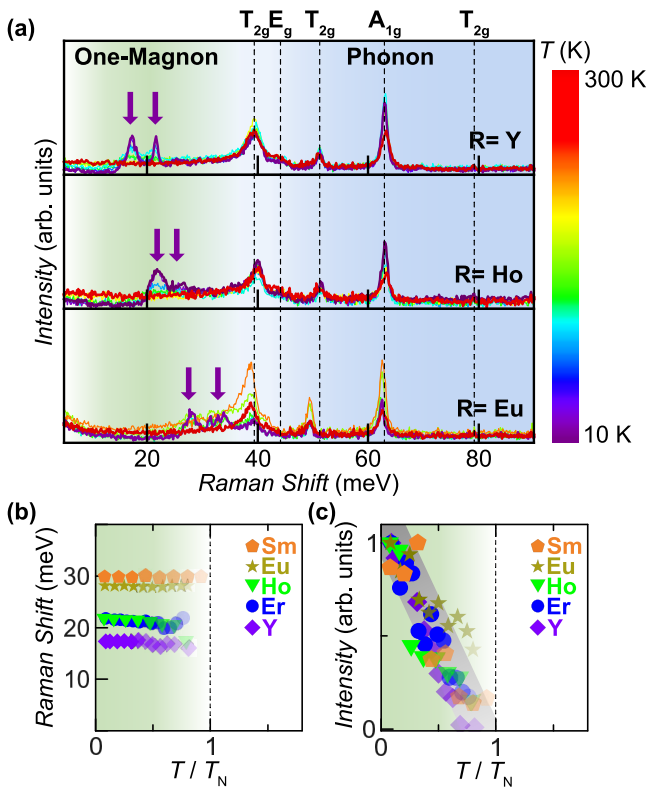}
    \caption{\textbf{Raman data of $A_2$Ru$_2$O$_7$}. (a) Temperature dependent Raman spectra of Y$_2$Ru$_2$O$_7$, Ho$_2$Ru$_2$O$_7$, and Eu$_2$Ru$_2$O$_7$, measured in the temperature range of 10 K to 300 K. Phonon modes between 40 - 70 meV are shaded in blue and their symmetries are assigned via vertical dotted lines. One-magnon modes between 15 - 35 meV are shaded in green. The purple arrows mark one-magnon modes below \textit{T}\textsubscript{N}. (b) Temperature dependence of the one-magnon energies for $A_2$Ru$_2$O$_7$. (c) Temperature dependence of the normalized one-magnon intensities. The thick line is a guide to the eyes. Dashed lines mark \textit{T}\textsubscript{N}.}
    \label{fig2}
\end{figure}

To extract information about the magnetic excitations, we will now focus on temperature-dependent Raman scattering results. From the six compounds that were synthesized and measured, we choose three compounds as an example to display in Fig.~\ref{fig2}(a). For all compounds a detailed temperature dependence was measured [see Supplemental Material Fig. S2~\cite{suppl}]. In the range of 40 meV – 65 meV, the spectra are dominated by phonon modes as marked by the blue shade in Fig. \ref{fig2}(a). Raman-active phonon modes obtained with the factor group analysis for the space group $Fd\bar{3}m$ ($A_{1g}$ + $E_g$ + $4T_{2g}$) were measured and confirmed for all compounds~\cite{NDSC}. The two sharp peaks around 50 meV and 60 meV are assigned as  $T_{2g}$ and $A_{1g}$ phonons, respectively~\cite{NDSC}. For the relatively broad peak near 40 meV, a partial overlap between $T_{2g}$ and $E_g$ modes is observed. Since the spectrum was taken in parallel polarization, the higher energy $E_g$ mode is weaker and tends to be absorbed by the stronger $T_{2g}$ channel, and for some compounds remains fully obscured. Of the remaining two  $T_{2g}$ modes, one is faintly observed near 80 meV while the other seems completely hidden either due to overlap with other modes or weak scattering intensity. The detailed assignment of the phonon peaks is given in Fig.~\ref{fig2}(a). No dramatic temperature-dependence is seen for the phonon modes indicating no sign of a structural phase transition. 

At energies below 40 meV, Raman spectra at T$>$\textit{T}\textsubscript{N} show a subtle enhanced background indicating paramagnetic fluctuations~\cite{EIOmagnon}. Below \textit{T}\textsubscript{N}, this scattering contribution is quenched and gives rise to well-defined, new excitations in the range of 15 - 35 meV [the purple arrows in Fig.~\ref{fig2}(a)]. This strongly suggests that these peaks are related to magnetic order. As we outline below, these modes can be assigned as one-magnon excitations. In Figs.~\ref{fig2}(b) and \ref{fig2}(c), the temperature dependent behavior of the energy and normalized intensity is plotted. The frequencies of the magnetic modes show very little thermal evolution and vanish for T$>$\textit{T}\textsubscript{N} as shown in Fig.~\ref{fig2}(b). On the contrary, the intensity is strongly temperature dependent: at the base temperature of 10 K it is the strongest, followed by an almost monotonic decrease as the temperature approaches \textit{T}\textsubscript{N}. In addition, the linewidth increases sharply close to \textit{T}\textsubscript{N} for all compounds [see Supplemental Material Fig. S3~\cite{suppl}]. All these observations are consistent with our interpretation as one-magnon modes. We note that sizable excitation energies of the magnon modes present in related pyrochlore magnets can render quantum and thermal fluctuations ineffective. Therefore, a significant softening in magnon energy may be lacking upon approaching \textit{T}\textsubscript{N} [as seen in Fig. \ref{fig2}(b)]~\cite{YIORaman,EIOmagnon, NDSC}.

\begin{figure*}
    \includegraphics[]{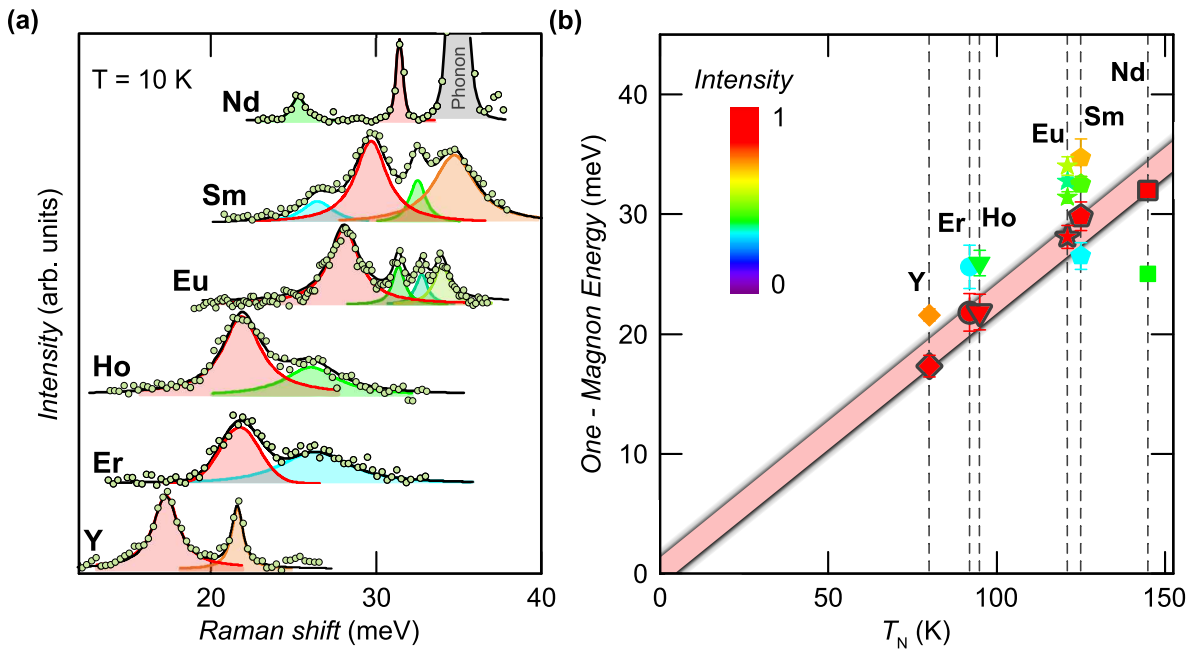}
    \caption{\textbf{Correlation between \textit{T}\textsubscript{N} and one-magnon excitation energies of $A_2$Ru$_2$O$_7$}. (a). Background subtracted low-energy spectral range for $A$$_2$Ru$_2$O$_7$ covering one-magnon modes measured at T = 10 K (full circles). The one-magnon excitations are shaded in colors respective to their relative intensity. Note that the spectrum of $A$ = Nd is obtained via measuring single crystals and the gray shaded peak near 38 meV stands for the $T_{2g}$ phonon~\cite{NDSC}. (b) One–magnon excitation energies measured at 10 K plotted with respect to \textit{T}\textsubscript{N} for each compound. Colors of the squares represent the normalized intensity of each peak (the strongest colored in shaded red). The vertical bars on the symbols correspond to the relative linewidths of the modes. Modes of strongest intensity are connected by a straight red line.}
    \label{fig3}
\end{figure*}

We can first rule out a phononic origin of the new excitations observed by Raman spectroscopy below \textit{T}\textsubscript{N}. Structural transitions often accompany magnetic transitions at \textit{T}\textsubscript{N}. This induces new Raman active phonon modes. However, there is no evidence for structural phase transition in these compounds~\cite{mag15Nd, DuijnPr227,mag13Eu,mag17Nd,mag9Ho,mag6Er}, nor would we expect phonon energies to be subject to such strong $A$-site ion dependence: While energies of the new excitations linearly scale with \textit{T}\textsubscript{N} [see Fig.~\ref{fig3}(b)], all phonon mode energies of our investigated compounds remain relatively unaffected by the choice of $A$-site ions [see Fig.~\ref{fig2}(a)].

Regardless, one could still be tempted to assign the origin of these new low energy excitations as split phonons of the doubly degenerate $E_g$ mode due a lowering of symmetry. However, the energy difference ($\Delta E$) between such two split phonon branches in pyrochlore lattice is typically an order of magnitude smaller ($\Delta E$ $\lessapprox$ 1-2 meV)~\cite{phononsplit1,phononsplit3,phononsplit2} than the gap between phonons and the new excitations observed in our system ($\Delta E$ $\approx$ 10-20 meV). Moreover, such a splitting is commonly accompanied by a dramatic renormalization of other phonon energies, which is absent in our data.

Excluding phonons, it is natural to consider the possibility for RE crystal field excitations. It is true that pyrochlore systems with heavy RE elements generally host excitations related to crystal field splitting. The energy scale of the RE crystal field energy levels is known to be comparable with that of the one-magnon modes in our study~\cite{RECFS1,RECFS2,RECFS3,RECFS4}. Moreover, inelastic neutron scattering studies conducted on Ho$_2$Ru$_2$O$_7$~\cite{mag8Ho,mag9Ho} uncovered crystal field transitions near 20 meV, which is in the proximity of the one-magnon modes from our Raman spectroscopic result.

However, as these excitations originate from the $A$-site atoms, there is no obvious reason why they should mimic the thermal evolution of the Ru spins. Indeed, as can be seen in Ref.~\cite{mag9Ho}, the intensity evolution of the crystal field excitations in Ho$_2$Ru$_2$O$_7$ with temperature is in stark contrast with that of the one-magnon modes in our Raman study: while the crystal field transition energy undergoes a slight energy shift at \textit{T}\textsubscript{N}, the intensity continuously \textit{increases} with increasing temperature~\cite{mag9Ho}. Meanwhile excitations reported in our study decrease in intensity with increasing temperature and vanish around \textit{T}\textsubscript{N}. Thus, while the RE crystal field splitting does seem to fall into a similar energy scale with the new low-energy peaks, their temperature behavior is in contradiction with what we observed, which strongly suggests that the low-energy peaks cannot be of crystal field origin.

In addition to the difference in thermal behavior, the linear correlation between the modes’ energies and \textit{T}\textsubscript{N}s of the respective compounds strongly suggests that they should be related to magnetism of the Ru atoms rather than to any $A$-site degree of freedom.

Finally, we note that in the case of Y$_2$Ru$_2$O$_7$, the Y\textsuperscript{3+} ions do not have any $4f$ electrons unlike the other RE compounds in this study. In the absence of $f$ electrons any crystal field splitting would arise from the $d$ electrons, which should occupy a much higher energy range (well above 1 eV). Yet, we observe the emergence of new low-energy modes below \textit{T}\textsubscript{N} in Y$_2$Ru$_2$O$_7$ as well, which follow the scaling behavior with \textit{T}\textsubscript{N}.

Ruling out phonons and RE crystal field excitations, only a magnetic origin remains since these materials are all insulators with about 200 - 600 meV charge gaps~\cite{tokura}. We can then consider two possibilities: two-magnon or one magnon modes.

Two-magnon Raman scattering processes are dominantly observed in magnetically ordered systems, whereas sharp one-magnon modes have vanishingly small scattering intensities~\cite{LF, RamanMag1, RamanMag8}. On the other hand, if the spins are coupled to other degrees of freedom (e.g., via spin-orbit coupling), the scattering intensity of one-magnon modes may be significantly enhanced~\cite{Ruclmagnon0,RamanMag14}. Indeed, in related systems hosting corner sharing $B$O$_6$ octahedra such as pyrochlore - $A$$_2$Ir$_2$O$_7$ ($A$=Y, Nd, Sm,  Eu), Nd$_2$Ru$_2$O$_7$~\cite{YIORaman,EIOmagnon,NDSC,RECFS4} - or perovskites - Sr$_2$IrO$_4$, Ca$_2$RuO$_4$~\cite{SIOmagnon1,SIOmagnon2, RamanMag9} - such one-magnon modes are prominently observed via Raman scattering at comparable energies. 

Next, we compare the characteristics of such enhanced one-magnons with those of typical two-magnon modes. Considering the rather narrow linewidth of our new magnetic scattering peaks [see Supplemental Material Fig. S3~\cite{suppl}], we may rule out two-magnon scattering as the origin, as such higher-order scattering processes lead to substantially enhanced linewidths and less well-defined lineshapes~\cite{RamanMag8}. Additionally, in contrast to a one-magnon process, two-magnon scattering oftentimes retains a finite intensity at and even beyond \textit{T}\textsubscript{N}, which reflects the existence of short-range correlations~\cite{twomagnon1,twomagnon2}. Finally, we note that two-magnon scattering in related pyrochlore systems was reported at significantly larger energies~\cite{RamanMag8}. Thus, we can confidently assign the low-energy peaks as one-magnon modes by ruling out other possible origins of the new Raman peaks below \textit{T}\textsubscript{N}.

Taking a closer look at the one-magnon peaks, we find one dominant peak and several side peaks separated by about 5 meV, while their energies and number of peaks differ between compounds depending on the $A$-site RE ions. Fig. \ref{fig3}(a) displays the background subtracted low-energy Raman spectra of $A$$_2$Ru$_2$O$_7$ ($A$= Y, Nd, Sm, Eu, Ho, Eu) at T = 10 K (the spectrum of Nd$_2$Ru$_2$O$_7$ is obtained from single crystals and is thoroughly discussed in a separate paper~\cite{NDSC}). The one-magnon modes are well-fitted with Lorentzian curves as shown in Fig. \ref{fig3}(a). Colors of the shade are matched with the relative intensity of the signals compared to the maximum intensity peaks of each respective compounds. Plotting the energy of all the one-magnon peaks measured at the base temperature 10 K vs. \textit{T}\textsubscript{N} reveals a close correlation [see Fig. \ref{fig3}(b)]. The colors of the symbols represent the relative intensity of the magnon peaks. By connecting the red colored points (i.e., the modes that dominate in intensity for each compound) and continuing the trend down to zero energy, we see that their magnon peak energy is directly proportional to \textit{T}\textsubscript{N}. 

What are the implications of this remarkable discovery of linear scaling between the energy of the dominant Raman-active magnon modes and \textit{T}\textsubscript{N}s in pyrochlore ruthenates? Note that different spin structures may display different magnon spectra and cannot necessarily be linearly scaled by \textit{T}\textsubscript{N}, just as different crystal structures are characterized by their unique phonon spectra. This suggests the possibility of a similar (or common) structure of Ru spins among all compounds, albeit the fact that Raman spectroscopy cannot measure the spin structure directly. A generic S=1 spin model for pyrochlores predicted various spin structures depending on parameters such as J, D and single-ion anisotropy energy (D$_z$)~\cite{Gao}. Different magnetic phases should provide different magnon spectra. Naturally this leads one to expect Raman spectra to show different characteristics for different J, D and D$_z$ which is largely affected by the $A$-site ions. Yet, we see dominant Raman magnon modes proportionally scaling with \textit{T}\textsubscript{N} and also the side peaks existing within about 5 meV higher energy almost for all compounds measured. The absence of the upper side peak in Nd$_2$Ru$_2$O$_7$ may be rationalized due to its energetic overlap with phonons [see Fig.~\ref{fig3}(a) and Supplemental Material Fig. S4~\cite{suppl}], although one or two peaks can be seemingly identified at higher energy besides the dominant peak as well. 

Hence, the observed proportionality between the energy of the dominant magnon mode and \textit{T}\textsubscript{N} should be suitably explained, if the compounds have similar or same spin structures. Neutron scattering experiments and M\"{o}ssbauer spectroscopy measurements report different Ru spin phases for different pyrochlore ruthenate compounds~\cite{Gao,Rev1,mag3Er,mag6Er,mag8Ho,mag9Ho,Gd227,mag11Y,mag14Nd,mag15Nd,mag17Nd,mag17Nd2}. Considering the phase diagram given by the spin-only S=1 Hamiltonian in Ref~\cite{Gao}, Y$_2$Ru$_2$O$_7$ and Pr$_2$Ru$_2$O$_7$ can be assigned as coplanar XY AFM 1 model ~\cite{Gao,DuijnPr227}. Extensive neutron scattering data were collected for both Ho$_2$Ru$_2$O$_7$ and Er$_2$Ru$_2$O$_7$~\cite{mag8Ho,mag9Ho,mag3Er,mag6Er}. Although Ho$_2$Ru$_2$O$_7$ is classified to be a splayed FM~\cite{mag8Ho,mag9Ho}, for Er$_2$Ru$_2$O$_7$, results indicated a collinear state~\cite{mag3Er,mag6Er} while the possibility of it existing in the non-coplanar XY AFM or coplanar XY AFM2 region was also given ~\cite{Gao}. The Ru spin structure in Nd$_2$Ru$_2$O$_7$ was assigned as either non-coplanar XY AFM or coplanar XY AFM 2 by recent neutron scattering measurement, similar to Er$_2$Ru$_2$O$_7$~\cite{ChoiNd227}. In short, different neutron scattering studies conducted on several compounds reported different spin structures depending on the $A$-site ion, while even for individual compounds, such as Er$_2$Ru$_2$O$_7$ and Nd$_2$Ru$_2$O$_7$, the neutron scattering results are still far from conclusive.

Next we discuss the side peaks other than the dominant peaks. As noted above, the number of magnon modes are different for all compounds. Interestingly, a side peak seen in Y$_2$Ru$_2$O$_7$ at about 22 meV is singular and distinctly sharp, while side peaks from other compounds are either multiply split or much broader as shown in Fig.~\ref{fig3}(a). We realize that there exists a tendency for compounds with a strong effective magnetic moment $\mu_{\mathrm{cal}}$ = g$_J$ $\sqrt{J(J+1)}$ ($A$=Er and Ho) to display broader and ambiguous side peaks. Those that have weaker magnetic moments ($A$=Sm and Eu) showed relatively distinct multiple sharp peaks. 

Peak splitting or broadening of the side peaks can be explained by the $A$-$B$ site interaction. It is known that the inter-sublattice interaction is critical in understanding the $A$-site magnetic structure. From previous works on pyrochlore ruthenates, for compounds with Ho, Er, and Gd at the $A$-site, it is known that the Ru ordering below \textit{T}\textsubscript{N} has a direct ordering effect on the $A$-site ions~\cite{Rev1,Gd227,mag8Ho,mag9Ho,mag3Er,mag6Er}. In line with this consensus, we apply the same magnetic mechanism to our Raman spectra. Since $A$ and $B$ site spins are mediated mainly by $A$-$B$ site exchange interaction at intermediate temperatures below \textit{T}\textsubscript{N} and above the low $A$-site ordering temperatures, it is natural to assume the side peaks from the Raman spectra also reflect these interactions. Considering that the trivalent $A$=Y ion is nonmagnetic, we must consider the two sharp magnon peaks of Y$_2$Ru$_2$O$_7$ as pure magnetic spin wave excitations of the $B$-site Ru ions. Additional side peaks observed from other compounds in the series instead directly reflect the $A$-$B$ site exchange interaction. The energy scale of few meVs apparent in the peak splitting and broadening is also consistent with the $A$-$B$ site exchange interaction. 
\section{Summary}
\label{Summary}
We present a systematic study on a series of pyrochlore ruthenates via temperature dependent Raman spectroscopy. Apart from the Raman-active phonon modes, new excitations that appear below \textit{T}\textsubscript{N} are assigned as one-magnon modes in the energy range of 15 - 35 meV. All features observed in our Raman spectra in that energy range and at low temperatures (well below \textit{T}\textsubscript{N}) show a common behavior regardless of the  $A$-site ion. In addition to a dominant peak, higher energy side peaks exist within the range of about 5 meV. These side peaks are multiply split or appear very broad, while only a single sharp peak appears for Y$_2$Ru$_2$O$_7$. As Y$_2$Ru$_2$O$_7$ is the only compound in our study without 4$f$ electrons, the observed peak broadening and splitting of the side peaks in other compounds results from $A$-$B$ site interaction, which has an energy scale comparable to that of the splitting. Remarkably, the dominant peaks show proportional behavior with the \textit{T}\textsubscript{N} values. Our Raman spectroscopy results imply a common structure of Ru spins for all investigated pyrochlore ruthenate compounds and open up the possibility of magnetic band engineering by mixing $A$-site ions. 
\begin{acknowledgments}
We thank Heung-Sik Kim and Sungkyun Choi for their fruitful discussion. This work was supported by the Institute for Basic Science (Grant No. IBS-R009-G2, No. IBS-R009-Y3) and National Research Foundation of Korea (NRF) grant funded by the Korea government (MSIT, Grant No. 2020R1A2C1011439, No. 2022R1A3B1077234). 
\end{acknowledgments}

\end{document}